\documentclass[letterpaper, 10 pt, conference]{ieeeconf} 
               
\IEEEoverridecommandlockouts                         
\let\labelindent\relax

\usepackage{amsmath,graphicx,psfrag,epsfig,color,amsfonts,hyperref}
\usepackage{enumitem}
\usepackage{version}
\usepackage{subfigure}
\usepackage{breqn}
\usepackage{mathtools,amssymb,lipsum,mathrsfs}

\usepackage{bibentry}

\newcommand\oprocendsymbol{\hbox{$\blacksquare$}}
\newcommand\oprocend{\relax\ifmmode\else\unskip\hfill\fi\oprocendsymbol}

\renewcommand{\theenumi}{\roman{enumi}}

\newtheorem{theorem}{Theorem}

\newtheorem{assumption}{Assumption}
\newtheorem{remark}{Remark}
\newtheorem{corollary}{Corollary}

\newtheorem{definition}{Definition}

\newcommand{\RR}{\mathbb{R}}

\title{\LARGE \bf

Global synchronization analysis of non-diffusively coupled networks through Contraction Theory
}

\author{Fatou K. Ndow$^{1}$ and Zahra Aminzare$^{2}$ 
\thanks{ This work is supported by Simon Foundations' grant $712522$ and NSF grant IOS$-2037828$.}
\thanks{$^{1}$Fatou K. Ndow is a Ph.D. candidate at the Department of Mathematics, University of Iowa, Iowa City, IA 52242, USA. 
        {\tt\small fatou-ndow@uiowa.edu}}%
\thanks{$^{2}$Zahra Aminzare is an Assistant Professor at the Department of Mathematics, University of Iowa, Iowa City, IA 52242, USA. 
        {\tt\small zahra-aminzare@uiowa.edu}}%
}

\begin{document}

\maketitle
\thispagestyle{empty}
\pagestyle{empty}


\begin{abstract}
Synchronization of coupled dynamical systems is a widespread phenomenon in both biological and engineered networks, and understanding this behavior is crucial for controlling such systems. Considerable research has been dedicated to identifying the conditions that promote synchronization in \textit{diffusively} coupled systems, where coupling relies on the difference between the states of neighboring systems and vanishes on the synchronization manifold. In particular, contraction theory provides an elegant method for analyzing synchronization patterns in diffusively coupled networks. However, these approaches do not fully explain the emergence of synchronization behavior in \textit{non-diffusively} coupled networks where the coupling does not vanish on the synchronization manifold and hence the dynamics on the synchronization manifold differ from the uncoupled systems. Inspired by neuronal networks connected via non-diffusive chemical synapses, we extend contraction theory to establish sufficient conditions for global synchronization in general non-diffusively coupled nonlinear networks. We demonstrate the theoretical results on a network of Hindmarsh-Rose oscillators connected via chemical synapses and networks of FitzHugh-Nagumo oscillators connected via chemical synapses and additive coupling.
\end{abstract}


\section{Introduction}\label{sec:Introduction}
The primary objective of this work is to investigate how neuron coupling affects the existence, stability, and robustness of synchronization patterns in neuronal networks. Neuron communication can occur through electrical or chemical transmissions, with one common form of electrical transmission being known as a \textit{gap junction}, which can be modeled by the voltage difference between the coupling neurons. This type of coupling is \textit{diffusive} and only relies on the difference between neighboring system states, which vanishes on the synchronization manifold where all states become identical. However, the diffusive coupling is not commonly found in the brain. Instead, chemical transmissions in the form of \textit{synaptic} coupling and \textit{autapse} (self-connection) are prevalent. The influence of these coupling dynamics and their interactions on synchronization has been studied in \cite{Karami_etal_ephaptic_2021, Fan_etal_Autapses_2018, Protachevicz_etal_Autapses_2020,Koppel_Ermentrout}, where they are modeled by the product of post-synaptic voltage and pre-synaptic synapse, resulting in \textit{non-diffusive} coupling. Unlike diffusive coupling, non-diffusive coupling does not vanish on the synchronization manifold.
\medskip 
 
The study of network synchronization in diffusively coupled networks using contraction theory \cite{Slotine_Wang_2005, Pham:Slotine:2007} and master stability function \cite{pecora_master_1998,Soriano2016Nonlinear} is a well-established research area. Complete synchronization in networks of homogeneous systems with diffusive coupling has been extensively explored in \cite{arcak2011_Automatica, aminzare_synchronization_2014, sj-pcv-fb:19q}. Cluster synchronization solutions in heterogeneous systems with diffusive coupling have been studied in \cite{2018_Aminzare_Dey_Davison_Leonard}, while stochastic synchronization in stochastic networks has been reported in \cite{Russo-Wirth-Shorten-2019, Russo-Shorten-2018, ASri22}. However, relatively little effort has been devoted to studying synchronization patterns that arise from non-diffusive coupling networks coupled through digraphs (directed graphs). This paper aims to address this gap in the literature.
\medskip 

This paper focuses on analyzing a network of homogeneous (identical) non-linear systems that  are interconnected through digraphs and non-diffusive coupling. Our objective is to establish conditions that ensure the existence of complete synchronization solutions and determine their global stability.

The paper is structured as follows. Section~\ref{sec:Preliminaries} presents the networks we are investigating and provides a review of the relevant concepts from contraction theory. Section~\ref{sec:results} presents our main result. In Section~\ref{sec:simulations}, we apply our theory to several neuronal networks and present numerical simulations. Finally, we conclude in Section~\ref{sec:conclusion}.

\section{Preliminaries}\label{sec:Preliminaries}
In this paper, we consider a network of $N$ coupled systems described by the following equations. For $i=1,\ldots,N,$
\begin{equation}\label{eq:main:network}
    \dot{X}_i= F(X_i)+\sigma\sum_{j=1}^{N} \alpha_{ij}{H}(X_i,X_j).
\end{equation}
\textbf{Intrinsic dynamics.} Each individual in the network is described by a state $X_i$ which is a vector function from $[0,\infty)$ to $\mathbb{R}^n$. 
The intrinsic dynamic of each node is modeled by $F\colon C \subset \RR^n\to \RR^n$, where $C$ is a  convex, open, and connected of $\RR^n$.

 \textbf{Coupling Dynamics.} The function $H\colon C \times C \to \mathbb{R}^n$  describes the connection between the nodes  and the positive constant $\sigma$ represents a uniform coupling strength across the network. The coupling function ${H}$ can either be 
\textit{diffusive},  
in which, $H(X_j,X_i)=0$ when $X_j=X_i$, 
or \textit{non-diffusive}, 
that is, $H(X_j,X_i)\neq0$ when $X_j=X_i$. 
Note that we are interested in non-diffusive coupling in this work.

\textbf{Network Topology.}  A weighted 
graph $\mathcal{G}$, with adjacency (or connectivity) matrix  $\mathcal{A} = [\alpha_{ij}]$, represents the network topology, i.e., $\alpha_{ij}>0$ if node $j$ is connected to node $i$, and zero otherwise. 
Note that $\mathcal{A}$ is not necessarily symmetric. That is, the underlying graph is a \textit{digraph}, i.e., a directed graph. Also, note that we allow weights and loops (self-connections) on digraphs. 
Let $d_i=\sum_{j} \alpha_{ij}$ be the in-degree of node $i$, and $\mathcal{D}$ be a diagonal matrix with $d_i$s on its diagonal. Then, $\mathcal{L}_{in} = \mathcal D - \mathcal A$ is the corresponding in-degree \textit{Laplacian} matrix with (at least) a zero eigenvalue and corresponding right eigenvector $\mathbf{1}=(1,\ldots,1)^\top.$ For the ease of notation, we drop \textit{in} from $\mathcal{L}_{in}$ and simply show it with  $\mathcal{L}$.  We denote the eigenvalues of $\mathcal{L}$ by $\lambda_{i,\mathcal{L}}$. Since $\mathcal{A}$ is asymmetric,  $\mathcal{L}$ is also asymmetric, and the eigenvalues might be complex. In addition, the eigenvectors do not form a basis for $\RR^n$. 
\medskip 

\textbf{A necessary condition for the existence of synchronization solutions.} 
Eq.~\eqref{eq:main:network} \textit{completely synchronizes} if for any solution $X$, there exists a solution $\psi= (X_s,\ldots,X_s)^\top$  such that $\|X(t)-\psi(t)\|\leq e^{ct} \|X(0)-\psi(0)\|$, for some synchronization rate $c>0$. 
A synchronization solution of the form $(X_s,\ldots,X_s)^\top$ exists only if $\sum_{j=1}^{N} \alpha_{ij}=k$, for all $i=1,\ldots,N$. That is, the adjacency matrix $\mathcal{A}$ possesses a constant row sum $k$. 
Under this condition, the synchronization solution $(X_s,\ldots,X_s)^\top$ satisfies
\begin{equation}\label{eq:main:syncSolution}
    \dot{X_s}= F(X_s)+\sigma k H(X_s,X_s).
\end{equation}

Note that for a diffusive coupling, the coupling term vanishes, i.e., $H(X_s, X_s)=0$. Therefore, the synchronization solution has the same dynamics as the isolated system. This is not true for a non-diffusive type of coupling, i.e., the synchronization solution may differ from the solutions of isolated systems, as shown in Eq.~\eqref{eq:main:syncSolution} and in Fig.~\ref{fig:fig1} and  Fig.~\ref{fig:fig2} in Sec.~\ref{sec:simulations} below.
\medskip 

\begin{assumption}\label{assm:necessarry}
For the rest of the paper, we assume that the connectivity matrix has a constant row sum equal to $k$ to ensure the existence of synchronization solutions for Eq.~\eqref{eq:main:network}.
\end{assumption}

\medskip 
\textbf{Main goal.} Assuming that the network possesses a synchronization solution, our main goal of this paper is to provide conditions that guarantee the global stability of the synchronization solution. Our condition is based on the intrinsic dynamics $F$, the coupling dynamics $\sigma$ and $H$, and the algebraic connectivity of the underlying graph. 
\medskip 

\textbf{Algebraic connectivity.} 
The Courant-Fischer Theorem states that the algebraic connectivity of an undirected graph, which measures its connectivity, is equivalent to $\lambda_{2,\mathcal{L}}$. However, for a digraph, this equivalence no longer holds. Therefore, to generalize the concept of algebraic connectivity to digraphs, we adopt the following definition from \cite[Section 2.3]{Chai2007}.

\medskip 

\begin{definition}\label{def:algebraic_connectivity}
For a digraph with Laplacian matrix $\mathcal{L}$, Fiedler algebraic connectivity
denoted by $\mathbf{a}(\mathcal{L})$,  is defined as follows
\begin{equation*}
\mathbf{a}(\mathcal{L}) = \min_{X\perp \mathbf{1}_N, X^\top X=1} X^\top \mathcal{L} X.
\end{equation*}
\end{definition}
 \medskip 

As mentioned above, for general digraphs, 
$\mathbf{a}(\mathcal{L})\neq \Re(\lambda_{2,\mathcal{L}})$. That is because the eigenvectors of an asymmetric Laplacian matrix do not form a basis of $\RR^n$ (required by the Courant-Fischer Theorem).
 Indeed, by \cite[Corollary 4.14]{Chai2007}, $\mathbf{a}(\mathcal{L})\leq \Re(\lambda_{2,\mathcal{L}})$ and the equality holds only when $\mathcal{L}$ is a normal matrix which is a rare occurrence in practical applications. 
Following \cite[Theorem 2.35]{Chai2007}, the algebraic connectivity $\mathbf{a}(\mathcal{L})$ is equal to the smallest eigenvalue of $Q^{\top}\frac{\mathcal{L}+\mathcal{L}^{\top}}{2}Q$ where $Q$ is an ${N\times (N-1)}$ matrix whose columns form an orthonormal basis for $\text{span}\{\mathbf{1}_N\}^{\perp}$.  
\medskip 

\begin{assumption}\label{assm:connectivity}
A digraph is connected if it contains  a node that is reachable from other nodes in the graph. 
For a connected digraph, $\mathbf{a}(\mathcal{L})>0$. In this work, we assume that the graphs are connected. 
\end{assumption}

\medskip 
\textbf{Contraction Theory and Logarithmic Norm.} To establish our condition, we will use contraction theory, which is an elegant tool for studying \textit{global} synchronization stability. 

Nonlinear system $\dot X = {F}(X)$ is  \emph{contractive}, if any two trajectories $X$ \& $Y$ converge to each other exponentially, that is, for some $c<0$,
\[\|X(t)-Y(t)\|\leq e^{c t}\|X(0)-Y(0)\|.\] 
A proper tool for characterizing contractive systems 
is provided by the \textit{logarithmic norms} 
of the Jacobian $DF$ of the
vector field, evaluated at all possible states (see \cite{Lewis1949, Hartman1961, Loh_Slo_98, Pav_Pog_Wou_Nij, Russo}). That is, 
if  for \textit{some} norm $\|\cdot\|_{\mathcal{X}}$ and some negative constant $c$, 
\begin{equation*}\label{eq:mu_negative}
\sup_{X}\mu_{\mathcal{X}}[DF(X)]\leq c <0, 
\end{equation*}
then $\dot X= F(X)$ is contracting. This framework analyzes a nonlinear system via an infinite family of local linearization. The key insight is that if all solutions are locally stable, then they are globally stable.

\section{Main Results}\label{sec:results}
We aim to establish conditions that foster synchronization in non-diffusively coupled networks of nonlinear systems. 
\medskip 

\begin{theorem}\label{thm:main_result}
Consider the network given in Eq.~\eqref{eq:main:network} with  Assumptions~\ref{assm:necessarry} and \ref{assm:connectivity}. Suppose that there exists  a positive definite matrix $P$ 
such that $P D_2H(X,X)+ D_2H(X,X)P$ is positive semidefinite, where $ D_2H$ is the derivative of $H$ with respect to the second argument and is assumed to be a non-negative diagonal matrix.  
Then, for any solution $X$ of Eq.~\eqref{eq:main:network}, there exists a solution $X_s$ of Eq.~\eqref{eq:main:syncSolution} 
 such that 
\begin{equation*}
    \|X(t)-\bar X(t)\|_{2,I_N\otimes\sqrt P }\;\;\leq\; e^{ct} \|X(0)-\bar X(0)\|_{2, I_N\otimes\sqrt P}
\end{equation*}
where $\bar X = (X_s,\ldots,X_s)^\top$,  
$
 c = \sup_{X\in C} \hskip 0.1cm \mu_{2,\sqrt P}[\mathbf{A}(X)]
$
and the matrix $\mathbf{A}(X)$ is equal to 
\[DF(X)+\sigma k(D_1H+D_2H)(X,X) -\sigma \mathbf{a}(\mathcal{L})D_2H(X,X).\]
Furthermore, the network  synchronizes when $c<0$.
\end{theorem}

\medskip 
\begin{proof}
Let $e_i:=X_i-X_s$, 
$e= (e_1,\ldots,e_N)^\top$, and
\[V(e):=\frac{1}{2}e^\top (\mathbf{I}_N \otimes P) e=\frac{1}{2}\sum_{i=1}^Ne_i^\top P e_i,\]
where $\otimes$ is the Kronecker product and $\mathbf{I}_N$ is an $N\times N$ identity matrix.
We need to show that $\dot V(e) \leq 2c V(e)$.
Since $X_i$ satisfies Eq.~\eqref{eq:main:network} and $X_s$ satisfies Eq.~\eqref{eq:main:syncSolution}, $\dot V$ becomes
\begin{align*}
    &\dot{V}(e) = \sum_{i=1}^{N}e_i^\top P(F(e_i+X_s)-F(X_s)) \\
& + \sigma\sum_{i=1}^{N}e_i^\top P\sum_{j=1}^N \alpha_{ij}(H(e_i+X_s,e_j+X_s)-H(X_s,X_s)). 
\end{align*}
Using Taylor expansions  for a single variable and multivariable vector-valued functions, respectively, for the first and second terms above, we get
\begin{align*}
\dot{V}&= \sum_{i=1}^{N}e_i^\top P DF(X_s) e_i\\
        &+\sigma\sum_{i,j=1}^N e_i^\top P \alpha_{ij}\left(D_1H(X_s,X_s)e_i+D_2H(X_s,X_s)e_j\right) \\
     &= e^{\top} (\mathbf{I}_N \otimes P)(\mathbf{I}_N \otimes  DF(X_s))e \\
     &+ \sigma ke^{\top} (\mathbf{I}_N \otimes P)(\mathbf{I}_N \otimes  D_1H(X_s,X_s))e \\
     &+\sigma e^{\top} (\mathbf{I}_N \otimes P)(\mathcal{A}\otimes  D_2H(X_s,X_s))e  
\end{align*}
 Since $\mathcal{L}=k \mathbf{I}_N-\mathcal{A}$, the last term of the above equation can be written in terms of the Laplacian matrix as follows. 
\begin{equation}\label{eq:Vdot}
\begin{split}
 \dot{V}(e) &= e^{\top} (\mathbf{I}_N \otimes  P DF)e + \sigma ke^{\top} (\mathbf{I}_N \otimes  P D_1H)e \\
           &+\sigma ke^{\top} (\mathbf{I}_N \otimes P D_2H)e -\sigma e^{\top} (\mathcal{L}\otimes  P D_2H)e. 
 \end{split}
\end{equation}
Since $P D_2H(X,X)+ D_2H(X,X)P$ is  positive semidefinite, there exists a symmetric matrix $M(X)$ such that  $P D_2H(X,X)+ D_2H(X,X)P = 2 M(X)^\top M(X)$. 
Hence, the last term of Eq.~\eqref{eq:Vdot} can be written as
\begin{equation*}
    \begin{split}
     e^{\top}(\mathcal{L}\otimes P D_2H)e &= \frac{1}{2} e^{\top}(\mathcal{L}\otimes (P D_2H+ D_2H P))e\\
     &= e^{\top}(\mathcal{L}\otimes (M^\top M))e\\
     &=  ((\mathbf{I}_N \otimes M)e)^\top(\mathcal{L}\otimes \mathbf{I}_m)(\mathbf{I}_N \otimes M)e\\
       &\geq  \mathbf{a}(\mathcal{L})((\mathbf{I}_N \otimes M)e)^\top(\mathbf{I}_N \otimes M)e.
    \end{split}
\end{equation*}
The first equality holds because both $P$ and $D_2H$ are symmetric. The last inequality holds by the definition of $\mathbf{a}(\mathcal{L})$ and that  $((\mathbf{I}_N \otimes M)e)^\top(\mathbf{1}_N \otimes \mathbf{I}_n)=0$.  
Therefore,
\begin{align*}
   -e^{\top}(\mathcal{L}\otimes P D_2H)e &\leq -\mathbf{a}(\mathcal{L}) ((\mathbf{I}_N \otimes M)e)^\top(\mathbf{I}_N \otimes M)e\\
   &=- \mathbf{a}(\mathcal{L})e^{\top}(\mathbf{I}_N\otimes (M^\top M))e\\
      &= -\mathbf{a}(\mathcal{L}) e^{\top} (\mathbf{I}_N \otimes PD_2H)e.
\end{align*}
Incorporating the above inequality into Eq.~\eqref{eq:Vdot}, we obtain
\begin{equation*}
    \begin{split}
 \dot{V}&(e) \leq e^{\top} (\mathbf{I}_N \otimes  P DF)e + \sigma ke^{\top} (\mathbf{I}_N \otimes  P D_1H)e \\
 &+\sigma ke^{\top} (\mathbf{I}_N \otimes P D_2H)e - \sigma \mathbf{a}(\mathcal{L})e^{\top} (\mathbf{I}_N\otimes  P D_2H)e \\
 &= \sum_{i=1}^{N}e_i^\top P [DF+\sigma k (D_{1}H+  D_{2}H)- \sigma \mathbf{a}(\mathcal{L})D_{2}H]e_i\\ 
  &\leq \frac{2 c}{2} \sum_{i=1}^{N}e_i^\top P e_i= 2c V(e). 
    \end{split}
\end{equation*}
The last inequality holds by the quad condition \cite{Delellis_DiBernardo_Russo_2011} and gives the desired result. 
\end{proof}

\medskip 
In what follows, we will examine specific instances of Theorem~\ref{thm:main_result} and compare them with previous studies.  

\begin{description}[leftmargin=*]
\item[Diffusive coupling] Eq.~\eqref{eq:main:network}  is diffusively coupled if the coupling vanishes on the synchronization manifold, that is, $H(X,X)=0$. In this case, the coupling function can be written as $H(X,Y)=D(Y-X)$ for some non-negative diagonal  matrix $D$, or more generally, $H(X,Y)=G(Y)-G(X)$ for some nonlinear function $G$ with non-negative diagonal $DG$. 
Therefore, $D_1H+D_2H=0$, and 
 $c$ in Theorem~\ref{thm:main_result} becomes
\begin{equation}\label{eq:diffusive}
c = \sup_{X\in C} \hskip 0.1cm \mu_{2,\sqrt{P}}[DF(X)-\sigma \mathbf{a}(\mathcal{L})DG]. 
\end{equation}
The classical condition for synchronization, as studied in \cite{arcak2011_Automatica}, can be represented by the synchronization rate $c$ when an undirected graph is used to model the network. In this case,  $\mathbf{a}(\mathcal{L})=\lambda_{2,\mathcal{L}}$. However, Eq.~\eqref{eq:diffusive} is valid for any networks coupled through non-linear diffusive coupling (as studied in, e.g., \cite{2008_Liu_Chen}) and arbitrary digraphs. 
\medskip 

\item[Additive coupling]  If $H(X,Y)$ only depends on $Y$, say $H(X,Y)=Y$, the coupling is called \textit{additive} which  is a common type of coupling among neurons. In this case, since $D_1H=0$, the coupling term helps $c$ being negative, only if $k-\mathbf{a}(\mathcal{L})<0$. An example of such a digraph is a complete graph ($\mathbf{a}(\mathcal{L}) = k+1$). 
An example of two coupled FitzHugh-Nagumo with this type of coupling is studied in \cite{Pham:Slotine:2007}. 
In Sec.~\ref{sec:simulations} below, we will study an arbitrary network of FN oscillators coupled through a digraph with $k-\mathbf{a}(\mathcal{L})<0$ and  additive coupling. 
Note that a network coupled through additive coupling with $k-\mathbf{a}(\mathcal{L})>0$ may still synchronize, but since our theory only provides sufficient conditions, it cannot justify  its synchronization. 
\medskip 

\item[Excitatory synaptic coupling] In an excitatory coupling, $D_1H$ is negative definite, so it helps $c$ to be more negative, and hence it facilitates network synchronization. On the other hand, since $D_2H$ is positive semidefinite, depending on the sign of $k-\mathbf{a}(\mathcal{L})$, $D_2H$ may or may not help to make $c<0$.  We will discuss two examples with excitatory synaptic coupling in Sec.~\ref{sec:simulations} below. 
\medskip 

\item[Balanced graphs] A graph is called  
\textit{balanced} if the in-degree of each node is equal to its out-degree.
For example, an undirected graph is a balanced graph. A graph is a \textit{k-regular} graph if each node has in- and out-degree equal to $k$.
Note that a k-regular graph is balanced, while a balanced graph is not necessarily a k-regular graph.
However, in this paper, according to Assumption \ref{assm:necessarry}, whenever we use a balanced graph in Eq.~\eqref{eq:main:network}, it must be a k-regular graph for some k. 
Suppose $\mathcal{G}$ is a balanced digraph with asymmetric in-degree Laplacian $\mathcal{L}$.  Then, $\mathbf{1}^\top\mathcal{L}=0$ in addition to $\mathcal{L}\mathbf{1}=0$. This means, $\mathcal{L}_{sym}=\frac{1}{2}(\mathcal{L}+\mathcal{L}^\top)$ which is the symmetric part of $\mathcal{L}$,  is also a Laplacian matrix with $\text{spec}(\mathcal{L}_{sym})$ equals to $0\leq \lambda_{2,\mathcal{L}_{sym}} \leq \cdots \leq \lambda_{N,\mathcal{L}_{sym}}$. Therefore, by Courant-Fischer Theorem, for a balanced graph, the synchronization rate $c$ in Theorem~\ref{thm:main_result} becomes 
\begin{equation*}
        c = \sup_{X\in C} \hskip 0.1cm \mu_{2,\sqrt P}[DF+\sigma k(D_1H+D_2H) -\sigma \lambda_{2,\mathcal{L}_{sym}}D_2H].
\end{equation*}
\end{description}

\textbf{Some remarks and future directions}
\begin{enumerate}[leftmargin=5.5mm]
\item Theorem~\ref{thm:main_result} assumes a constant weight matrix $P$, but it is possible to consider a state-dependent matrix $P(X)$. If we do so, the matrix $\mathbf{A}$ becomes
\[DF+\sigma k(D_1H+D_2H)-\sigma \mathbf{a}(\mathcal{L})D_2H+P^{-1}\dot P.\]
While we do not explore this case in the current paper, it is a topic that we plan to investigate in the future.

\item  Theorem~\ref{thm:main_result} is currently restricted to weighted $L^2$ norms. Since it may not always be practical to identify the appropriate weight for a given network, it is essential to extend this result to non-$L^2$ norms. However, this remains an open question for future research.

\item The network represented by Eq.~\ref{eq:main:network} exhibits full homogeneity, which implies that all nodes share the same intrinsic and coupling dynamics as well as inputs. As a consequence, complete synchronization is expected, as explained in Theorem~\ref{thm:main_result}. Nevertheless, real-world networks typically involve non-identical components. Thus, we aim to extend our theory and investigate the stability of \textit{cluster synchronization} solutions in heterogeneous networks, where heterogeneity may arise in the intrinsic dynamics $F$, the coupling $H$, or graph properties such as the in-degree $k$.
\end{enumerate}

\section{Application to neuronal networks and numerical simulations}\label{sec:simulations}
In this section, we apply our main result to two neuronal models that are connected non-diffusively. 

\subsection{\textbf{Networks of bursting Hindmarsh-Rose oscillators}}
\begin{figure}
    \centering
    \includegraphics[width=0.25\textwidth]{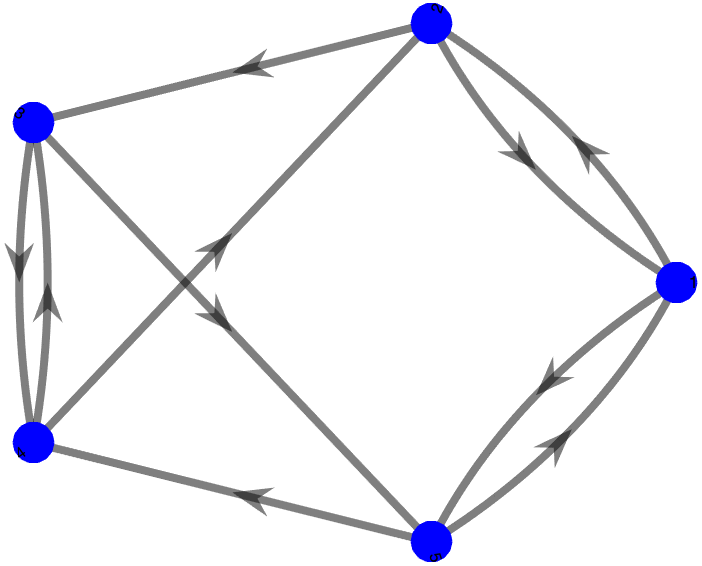}
    \bigskip 
    
\includegraphics[width=0.4\textwidth]{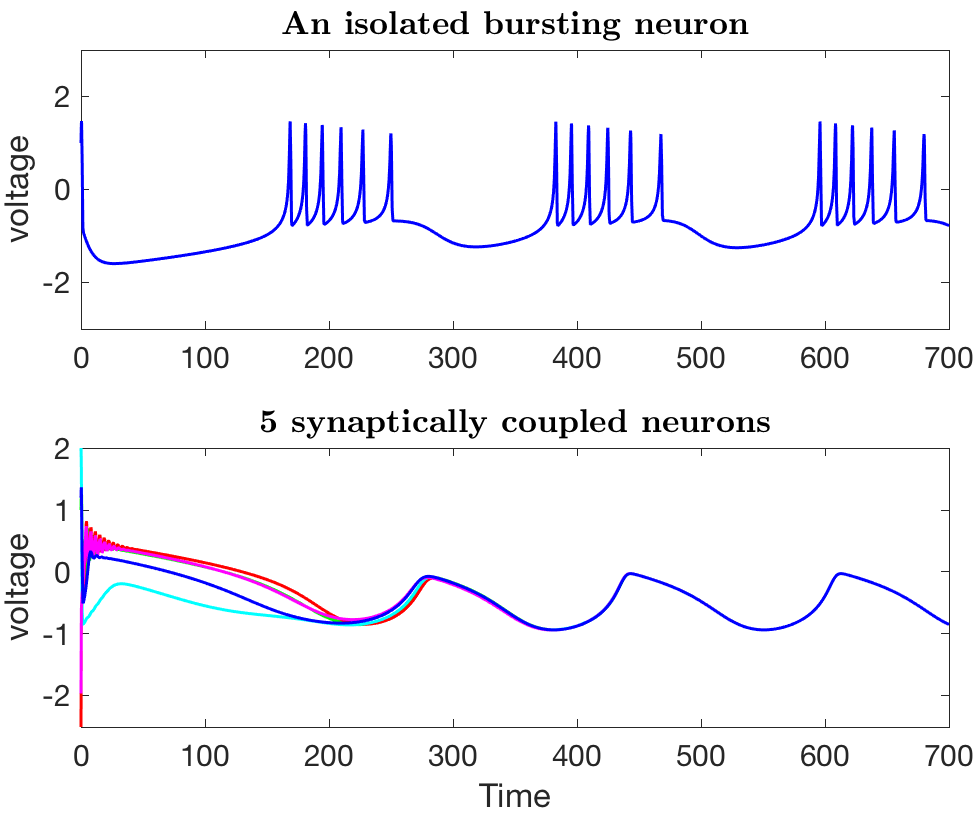}
\bigskip 

        \includegraphics[width=0.45\textwidth]{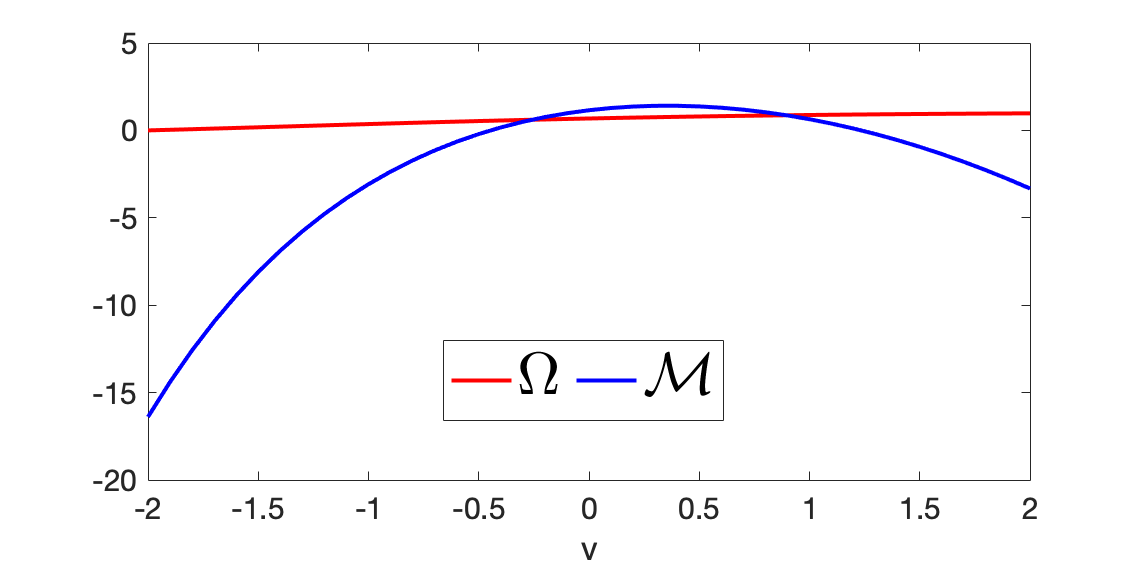}
        \caption{(Top) A digraph of five nodes with $k=2$ and and $\mathbf{a}_1(\mathcal{L})=1.382$ is shown. (Middle) The voltage variables of a single HR and five HR  that are synaptically coupled via the graph on the top panel are shown. (Bottom) The functions $\Omega(v)$  which is positive and $\mathcal{M}(v)$ which has a positive maximum are shown. These functions are defined in the proof of Corollary~\ref{cor:HR}. }
    \label{fig:fig1}
\end{figure}

The Hindmarsh-Rose (HR) model is a 3-dimensional system (a reduced model of the 4-dimensional Hodgkin-Huxley model) that describes how action potentials in neurons are initiated and propagated. The dynamics of a single HR model is described by $\dot{v}=\alpha v^2-v^3-w-n$, $\dot{w}=\beta v^2-w$, $\dot{n}=\epsilon(\gamma v+\delta-n)$ where $v$ is the voltage, and $w$ and $n$ are the gating variables related to the fast and slow currents, respectively \cite{HR_1984}. See Fig.~\ref{fig:fig1}(middle) for a numerical example of a bursting HR model, where at each cycle, a sequence of spikes is followed by a quiescent duration.  
In what follows, we consider a network of $N$ synaptically coupled HR models and study its synchronization behavior. For $i =1, \ldots, N$, the system
\begin{equation}\label{eq:HR:Network}
    \begin{split}
        \dot{v}_i &= {\alpha}v^2_i-v^3_i-w_i-n_i+\sigma(V_s-v_i)\sum_{j=1}^{N} \alpha_{ij} \Gamma{(v_j)}\\
        \dot{w}_i &= {\beta}v^2_i-w_i\\
        \dot{n}_i &={\epsilon}({\gamma}v_i+{\delta}-n_i),
    \end{split}
\end{equation}
describes the dynamics of each HR in the network. The sigmoidal function 
\[\Gamma{(v_j)}=\dfrac{1}{ 1+\exp(-\frac{2}{3}(v_j-\Theta_s))}\]
describes the synaptic input from neuron $j$ to neuron $i$, i.e., $H(X_i,X_j)=((V_s-v_i)\Gamma(v_j), 0, 0)^{\top}$. 
Here, the state is $X_i= (v_i,w_i,n_i)^\top$, and we assume $v_i$ belongs to the interval $[-1,1]$, which  contains the synchronization solutions.
\medskip 

\begin{corollary}\label{cor:HR}
Any networks of HR models described in Eq.~\eqref{eq:HR:Network} with $\sum_j \alpha_{ij}=k$ 
completely synchronize   if 
$\sigma >  \dfrac{\bar{\mathcal{M}}}{k},$
where $\bar{\mathcal{M}}$ is a positive constant that depends on the HR model parameters. 
\end{corollary}
\medskip 

\begin{proof}
We will show that for $P=\text{diag}\{1,{p^2},\frac{1}{{\epsilon} \gamma}\}$, 
\[c = \sup_{X\in C} \hskip 0.1cm \mu_{2,\sqrt P}[\mathbf{A}(X)]= \sup_{X\in C} \hskip 0.1cm\lambda_{max}[\mathbf{B}(X)] <0, \]
where $\mathbf{A}$ is as defined in Theorem~\ref{thm:main_result}, $\mathbf{B}$ is equal to 
\[\mathbf{B}=\sqrt{P}\;\frac{\mathbf{A}+(\mathbf{A})^\top}{2}(\sqrt{P})^{-1}=\left(\begin{array}{ccc} -A & B & 0 \\ B & -1 & 0 \\0 & 0 & -\epsilon\end{array}\right),\]
and $A,B$ and $p^2$ are as follows. 
\begin{align*}
A &=-2\alpha v+3v^2+\sigma k\Gamma(v)-\sigma (k-\mathbf{a}(\mathcal{L}))(V_s-v)\Gamma^{'}(v) \\
B &= \beta p v - \frac{1}{2p}, \quad p^2 =\dfrac{3}{\beta^2(1+(2\alpha-\beta)^2)}. 
\end{align*}
To show $c<0$, we only need to show that for any $v$ in the domain, $\text{Det}(\mathbf{B}) = A-B^2>0$ (this concludes that $A>0$ and hence the trace is negative). Since the graph is connected, the algebraic connectivity is positive. Therefore, $k-\mathbf{a}(\mathcal{L})<k$, and 
$k\Gamma(v)- (k-\mathbf{a}(\mathcal{L}))(V_s-v)\Gamma^{'}(v)<k\Omega(v)$, where
$$\Omega(v)= \Gamma(v)-(V_s-v)\Gamma^{'}(v).$$
For $\Gamma(v)$ that we chose here,  $\Omega(v)>0$ (see Fig.~\ref{fig:fig1}(bottom) for a numerical example and Remark\ref{remark:HR} below for more explanation). Therefore, $A-B^2>0$ if 
$\sigma >\dfrac{1}{k} \sup_v\mathcal{M}(v),$
where 
\[\mathcal{M}(v)= \frac{2\alpha v-3v^2 + B^2(v)}{\Omega(v)}.\]
Note that $\mathcal{M}$ is a continuous function on a compact set, so it admits a maximum, say $\bar{\mathcal{M}}.$ See Fig.~\ref{fig:fig1}(bottom) for a numerical example).
\end{proof}
\medskip 

\begin{remark}\label{remark:HR}
The sigmoidal function $\Gamma$ contains two parameters, $\eta:=-2/3$ and $\Theta_s$.  For $k-\mathbf{a}(\mathcal{L})<0$, e.g., a complete digraphs, for any $v\leq V_s$ and any $\eta$,  $k\Gamma(v)- (k-\mathbf{a}(\mathcal{L}))(V_s-v)\Gamma^{'}(v)>0$. However, for $k-\mathbf{a}(\mathcal{L})>0$, the choice of $\eta$ becomes important, that is, we need to choose $\eta$ small enough to decrease the effect of $\Gamma^{'}$ on the synchronization stability. This example has been studied in \cite{Belykh2015,Belykh_deLange_Hasler_2005} for normal digraph and relatively large $\eta$. For  $k$ large, they studied only the synchronization on the burst part of the oscillator. Here, we control the coupling function to reduce its effect when $k$ is large and study the synchronization on the whole part of the oscillator. 
\end{remark}
\medskip 

\textbf{Numerical Example} This is a numerical example to illustrate the result of Corollary~\ref{cor:HR}. 
Consider a network of five HR oscillators as described in Eq.~\eqref{eq:HR:Network} with parameters
$\alpha=2.8$, $\beta=4.4$,  $\delta=8$, 
$\gamma=9$, $\epsilon=1.6$, $V_s=1$, and $\Theta_s=-2$, that are coupled through a digraph with in-degree $k=2$, as shown in Fig.~\ref{fig:fig1}(top). 
For this example, $\bar{\mathcal{M}}\approx 1.408$ (the maximum of the blue curve in Fig.~\ref{fig:fig1}(bottom)) and $k=2>\mathbf{a}(\mathcal L)\approx1.382$. Therefore, our theory guarantees synchronization for $\sigma>1.408/2=0.704$. Fig.~\ref{fig:fig1}(middle) depicts the time series of the voltage variable for a single HR model and the synchronized voltages of the network of HR models.  Note the difference between the dynamics of uncoupled and coupled  oscillators due to non-diffusive coupling. 

\subsection{\textbf{Networks of  spiking FitzHugh-Nagumo Oscillators}}
The two dimensions FitzHugh-Nagumo (FN) model is a simplified version of the Hodgkin-Huxley model that can capture the essential properties of neuron action potentials. In this section, we explore a network of FN models that are coupled through two distinct mechanisms: chemical synapses, represented by a sigmoidal function, and additive coupling, modeled by a linear function. Our objective is to investigate the impact of these coupling mechanisms on the synchronization patterns of the network.

\subsubsection{Sigmoidal Coupling}
A network of $N$ synaptically coupled FN is described as follows. For  $i =1, \ldots, N$, 
\begin{equation}\label{eq:FN:Network}
    \begin{split}
         \dot{v}_{i} &= v_i-v_{i}^3/3-a-w_i+I+\sigma(V_s-v_{i})\sum_{j=1}^{N}\alpha_{ij}\Gamma(v_j)\\
     \dot{w}_{i} &=\epsilon(v_i-bw_i)
    \end{split}
\end{equation}
where $\Gamma$ is a sigmoidal function as follows 
\[\Gamma(v_j) = \dfrac{1}{1+\beta(1+\exp(-0.1(v_j-\Theta_s)))}\]
with  $\beta>0$ and an arbitrary threshold $\Theta_s$. 
Then, the coupling is $H(X_j,X_i)=((V_s-v_i)\Gamma(v_j), 0)^{\top}$ where the state is $X_i= (v_i,w_i)^\top.$ Note that we assume $V_s-v_i>0$ to ensure $D_2H$ is positive semidefinite.  Indeed, we assume $v_i\in[-2,2]$. 
\medskip 

\begin{corollary}
Any networks of FN models described in Eq.~\eqref{eq:FN:Network} with $\sum_j \alpha_{ij}=k$ completely synchronize if
 $\sigma >\frac{\bar{\mathcal{M}}}{k}$, 
 where $\bar{\mathcal{M}}>0$ is a constant and depends on the parameters. 
\end{corollary}
\medskip 

\begin{figure}
    \centering
    \includegraphics[width=0.25\textwidth]{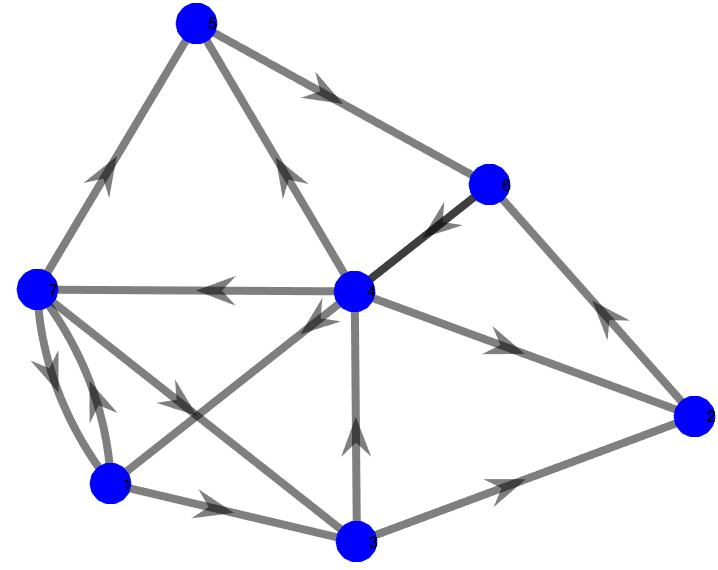}
    \bigskip 
    
    \includegraphics[width=0.4\textwidth]{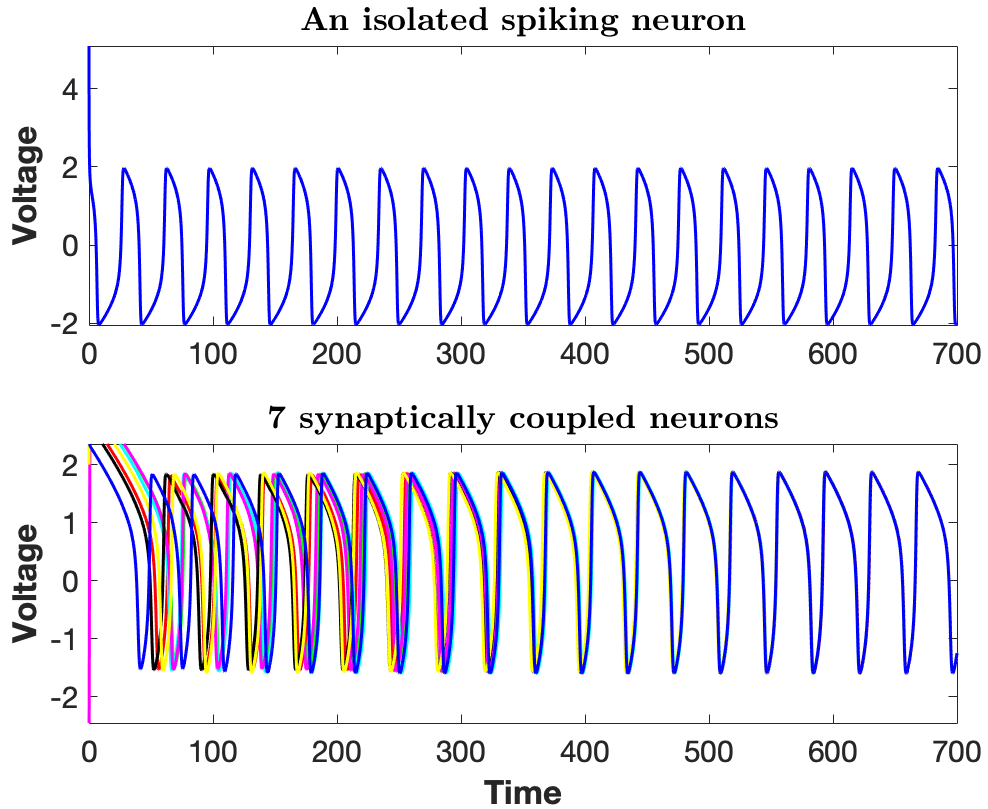}
         \bigskip 
         
    \includegraphics[width=0.45\textwidth]{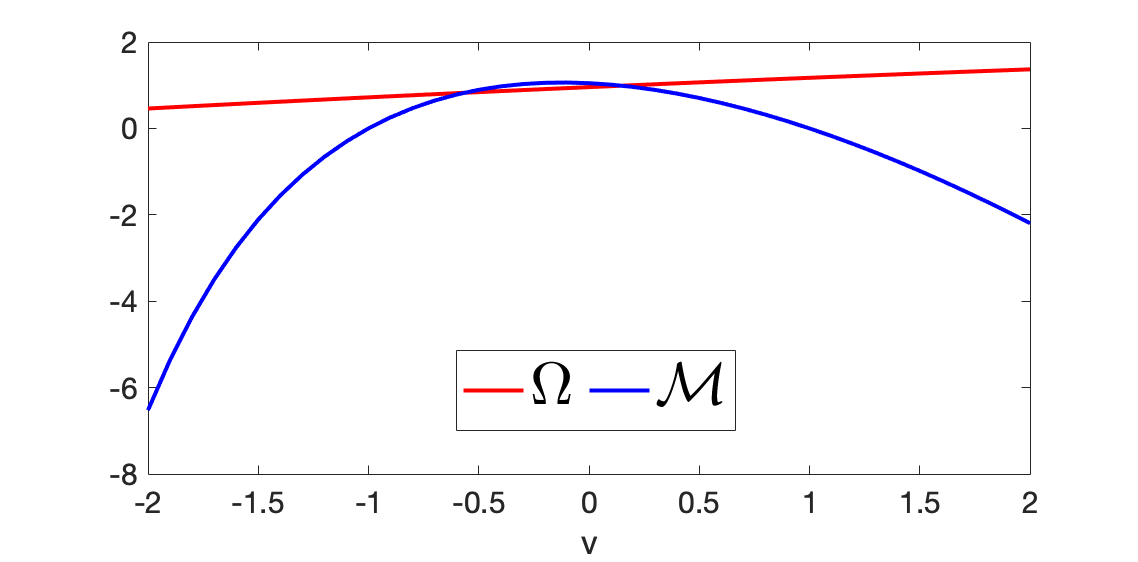}
    \caption{(Top) A digraph of seven nodes with $k=2$ and $\mathbf{a}_1(\mathcal{L})=1.3465$ is shown. (Middle) The voltage variables of a single FN and seven FN  that are synaptically coupled via the graph on the top panel are shown. (Bottom) $\Omega(v)$ in Eq.~\eqref{eq:Omega_v} which is positive and $\mathcal{M}(v)$ in Eq.~\eqref{eq:mathcal_M} which has a positive maximum are shown.}
    \label{fig:fig2} 
\end{figure}

\begin{proof}
We will apply Theorem~\ref{thm:main_result} by showing that for $P=\text{diag}\{1,\frac{1}{\epsilon}\}$, and $\sigma >\frac{\bar{\mathcal{M}}}{k}$ for some constant $\bar{\mathcal{M}}$ that we will determine later, 
\[c = \sup_{X\in C} \hskip 0.1cm \mu_{2,\sqrt P}[\mathbf{A}(X)]= \sup_{X\in C} \hskip 0.1cm\lambda_{max}[\mathbf{B}(X)] <0, \]
where $\mathbf{A}$ is as defined in Theorem~\ref{thm:main_result}, $\mathbf{B}$ is equal to 
\[\mathbf{B}=\sqrt{P}\;\frac{\mathbf{A}+(\mathbf{A})^\top}{2}(\sqrt{P})^{-1}=\left(\begin{array}{cc} A & 0   \\ 0  & -\epsilon b  \end{array}\right),\] 
 and
$ A =1-v^2-\sigma[ k\Gamma(v)-(k-\mathbf{a}(\mathcal{L}))(V_s-v)\Gamma^{'}(v)]. $
Obviously,   $c<0$ if $A<0$, or equivalently,  
\begin{equation*}\label{eq:sigma_FN}
    \sigma > \sup_v \frac{1-v^2}{k\Gamma(v)-(k-\mathbf{a}(\mathcal{L}))(V_s-v)\Gamma^{'}(v)}
\end{equation*}
Note that the denominator is always positive, so the inequality is valid. The reason is obvious when $k-\mathbf{a}(\mathcal{L})<0$. For  $k-\mathbf{a}(\mathcal{L})>0$, we use $k-\mathbf{a}(\mathcal{L})<k$ to argue that the denominator is bounded below by $k\Omega(v)$ where 
\begin{equation}\label{eq:Omega_v}
\Omega(v)=\Gamma(v)-(V_s-v)\Gamma^{'}(v)
\end{equation}
which is positive for FN as shown in Fig~\ref{fig:fig2}(bottom). 
Therefore, $\sigma >\frac{\bar{\mathcal{M}}}{k}$, where 
 $\bar{\mathcal{M}}= \sup_v \mathcal{M}(v)$, and 
 \begin{equation}\label{eq:mathcal_M}
\mathcal{M}(v)=\frac{1-v^2}{\Gamma(v)-(V_s-v)\Gamma^{'}(v)}.
 \end{equation}
  See Fig.~\ref{fig:fig2}(bottom) for a numerical example of $\mathcal{M}$. 
\end{proof}
\medskip 

\textbf{Numerical Example.}
Consider seven FN models as described in Eq.~\eqref{eq:HR:Network} with parameters 
$a=0.5$, $b=0.1$, $\epsilon=0.08$, $I=-2$, $V_s=35$, $\Theta_s=-20$ and $\beta=0.5$, that are coupled through a digraph with in-degree $k=2$, as shown in Fig.~\ref{fig:fig2}(top). 
For this example, $\bar{\mathcal{M}}\approx 1.06$ (the maximum of the blue curve in Fig.~\ref{fig:fig2}) and $k=2>\mathbf{a}(\mathcal L)\approx1.35$. Therefore, our theory guarantees synchronization for $\sigma>1.06/2=0.53$. Fig.~\ref{fig:fig2} depicts the time series of the voltage variable for a single FN model and the synchronized voltages of the network of FN models.
\medskip 

 \subsubsection{Additive Coupling}
 We now consider a network of  FN oscillators coupled through additive coupling (linear  non-diffusive functions) described by $H(X_i,X_j) = (v_j,0)^\top$ where $X_i=(v_i,w_i)^\top.$ 
  \medskip
  
\begin{corollary}
 Any networks of FN oscillators that are coupled through additive coupling and a digraph with $\sum_j \alpha_{ij}=k$ completely synchronize if $\sigma > \frac{1}{\mathbf{a}(\mathcal{L})-k}>0$. 
 \end{corollary}
 \medskip 
 
\begin{proof}
To apply Theorem~\ref{thm:main_result}, we let $P=\text{diag}\{1,\frac{1}{\epsilon}\}$ and show that for $\sigma > \frac{1}{\mathbf{a}(\mathcal{L})-k}>0$, 
\[c = \sup_{X\in C} \hskip 0.1cm \mu_{2,\sqrt P}[\mathbf{A}(X)]= \sup_{X\in C} \hskip 0.1cm\lambda_{max}[\mathbf{B}(X)] <0, \]
where $\mathbf{A}$ is as defined in Theorem~\ref{thm:main_result}, $\mathbf{B}$ is equal to 
\[\mathbf{B}=\sqrt{P}\;\frac{\mathbf{A}+(\mathbf{A})^\top}{2}(\sqrt{P})^{-1}=\left(\begin{array}{cc} A & 0   \\ 0  & -\epsilon b  \end{array}\right),\] and 
$A=1-v^2+\sigma(k-\mathbf{a}(\mathcal{L}))$. 
Note that here $D_1H=0$ and $D_2H=1$
To show $c<0$, we only need to show that $A<0$. Since $-v^2\leq0$, we  need to show that $1+\sigma(k-\mathbf{a}(\mathcal{L}))<0$, which is possible only if $k-\mathbf{a}(\mathcal{L})<0$ and $\sigma > \frac{1}{\mathbf{a}(\mathcal{L})-k}$. 
\end{proof}

\section{Conclusion}\label{sec:conclusion}

In this paper, we have expanded the application of contraction theory to determine the global stability of synchronization patterns arising from non-diffusively coupled networks, with a specific focus on neuronal networks. Our analysis provides both a necessary condition for the existence of synchronization solutions and a sufficient condition for their global stability. We applied our theory to two different neuronal models, the bursting Hindmarsh-Rose and spiking FitzHugh-Nagumo models, with two types of non-diffusive coupling: sigmoidal synapse and linear additive coupling. In all cases, we have shown that the bound for the coupling strength is inversely proportional to the digraphs' in-degree, which is consistent with results obtained in previous studies \cite{Pham:Slotine:2007, Belykh2015}. Our findings have important implications for understanding the behavior of complex networks, particularly those with non-diffusive coupling, and may have potential applications in various fields such as neuroscience, engineering, and ecology.




\end{document}